\documentclass[aps,prl,twocolumn,showpacs,superscriptaddress]{revtex4}

\pdfoutput=1
\usepackage{graphicx}
\usepackage{amsmath}
\usepackage{amssymb}
\usepackage{color}

\begin{document}

\title{Dynamical Phase Transitions and Instabilities in Open Atomic Many-Body Systems}

\author{Sebastian Diehl}
\affiliation{Institute for Theoretical Physics, University of Innsbruck, Technikerstr. 25, A-6020 Innsbruck, Austria}
\affiliation{Institute for Quantum Optics and Quantum Information of the Austrian Academy of Sciences, A-6020 Innsbruck, Austria}

\author{Andrea Tomadin}
\affiliation{Institute for Quantum Optics and Quantum Information of the Austrian Academy of Sciences, A-6020 Innsbruck, Austria}
\affiliation{ NEST, Scuola Normale Superiore and Istituto Nanoscienze - CNR, Pisa, Italy}

\author{Andrea Micheli}
\affiliation{Institute for Theoretical Physics, University of Innsbruck, Technikerstr. 25, A-6020 Innsbruck, Austria}
\affiliation{Institute for Quantum Optics and Quantum Information of the Austrian Academy of Sciences, A-6020 Innsbruck, Austria}

\author{Rosario Fazio}
\affiliation{ NEST, Scuola Normale Superiore and Istituto Nanoscienze - CNR, Pisa, Italy}

\author{Peter Zoller}
\affiliation{Institute for Theoretical Physics, University of Innsbruck, Technikerstr. 25, A-6020 Innsbruck, Austria}
\affiliation{Institute for Quantum Optics and Quantum Information of the Austrian Academy of Sciences, A-6020 Innsbruck, Austria}

\begin{abstract}
  We discuss an open driven-dissipative many-body system, in which the
  competition of unitary Hamiltonian and dissipative Liouvillian
  dynamics leads to a nonequilibrium phase transition. It shares
  features of a quantum phase transition in that it is interaction
  driven, and of a classical phase transition, in that the ordered
  phase is continuously connected to a thermal state. Within a
  generalized Gutzwiller approach which includes the description of
  mixed state density matrices, we characterize the complete phase
  diagram and the critical behavior at the phase transition approached
  as a function of time. We find a novel fluctuation induced dynamical
  instability, which occurs at long wavelength as a consequence of a
  subtle dissipative renormalization effect on the speed of sound.
\end{abstract}


\maketitle

Experiments with cold atoms provide a unique setting to study
nonequilibrium phenomena and dynamics, both in closed systems but also
for (driven) open quantum dynamics.  This relies on the ability to
control the many-body dynamics and to prepare initial states far from
the ground state. For closed systems we have seen a plethora of
studies of quench dynamics \cite{QuenchTh,QuenchExp}, thermalization
\cite{ThermalizationTh,ThermalizationExp}, and transport
\cite{Transport}, and also dynamical studies of crossing in a finite
time quantum critical points in the spirit of the Kibble-Zurek
mechanism \cite{KibbleZurekTh,KibbleZurekExp}.  On the other hand,
systems of cold atoms can be driven by external (light) fields and
coupled to dissipative baths, thus realizing driven open quantum
systems.  As familiar e.g. from the quantum optics of the laser, the
steady state of such a system (if it exists) is characterized by a
dynamical equilibrium between pumping and dissipation, and can exhibit
various nonequilibrium phases and phase transitions
\cite{ExcitonPolariton,Dalla09} as function of external control
parameters. In the present work we will study such scenarios for
quantum degenerate gases. Our emphasis is on understanding quantum
phases and dynamical phase transitions of cold atoms as an interacting
many-body condensed matter system far from equilibrium.

For a many-body system in thermodynamic equilibrium the competition of
two noncommuting parts of a microscopic Hamiltonian $H=H_{1}+gH_{2}$
manifests itself as a quantum phase transition (QPT), if the ground
states for $g\ll g_{c}$ and $g\gg g_{c}$ have different symmetries
\cite{SachdevBook}.  For temperature $T=0$ the critical value $g_{c}$
then separates two distinct quantum phases, while for finite
temperature this defines a quantum critical region around $g_{c}$ in a
$T$ vs.~$g$ phase diagram.  A seminal example in the context of cold
atoms in optical lattices is the superfluid--Mott insulator transition
in the Bose-Hubbard (BH) model, with Hamiltonian
\begin{equation}\label{eq:BHHamil}
H=-J\sum_{\langle \ell,\ell'\rangle}b_{\ell}^{\dagger}b_{\ell'} - \mu\sum_{\ell}\hat{n}_{\ell} +\tfrac{1}{2}U\sum_{\ell}\hat{n}_{\ell}(\hat{n}_{\ell}-1)~,
\end{equation}
with $b_{\ell}$ bosonic operators annihilating a particle on site
$\ell$, $\hat{n}_{\ell}=b_{\ell}^{\dagger}b_{\ell}$ number operators,
$J$ the hopping amplitude, and $U$ the onsite interaction strength.
For a given chemical potential $\mu$, chosen to fix a mean particle
density $n$, the critical coupling strength $g_c = (U/Jz)_{c}$
separates a superfluid $Jz \gg U$ from a Mott insulator regime $Jz \ll
U$ ($z$ the lattice coordination number).

In contrast, we consider a nonequilibrium situation in which the
competition of microscopic quantum mechanical operators results from
an interplay of unitary (Hamiltonian) and dissipative (Liouvillian)
dynamics.  We study a cold atom evolution described by a master
equation for the many-body density operator
\begin{eqnarray}
\partial_{t}\rho&=&  -i[H,\rho]+{\cal L}[\rho]~,\\\nonumber
\mathcal{L}[\rho] & = & \frac{1}{2}\kappa\sum_{\langle \ell,\ell' \rangle}\left(2 c_{\ell\ell'}\rho c_{\ell\ell'}^{\dagger}- c_{\ell\ell'}^{\dagger}c_{\ell\ell'}\rho- \rho c_{\ell\ell'}^{\dagger}c_{\ell\ell'}\right)~, 
\label{mastereq}
\end{eqnarray}
where
$c_{\ell\ell'}=(b_{\ell}^{\dagger}+b_{\ell'}^{\dagger})(b_{\ell}-b_{\ell'})$
are Lindblad ``jump operators'' acting on adjacent sites
$\langle\ell,\ell'\rangle$. The energy scale $\kappa$ is the
dissipative rate. As shown in \cite{Diehl08}, such dissipative
reservoir couplings are obtained in a setup where laser driven atoms
are coupled to a phonon bath provided by a second condensate.  For no
interaction $U=0$ this dissipation drives the system to a dynamical
equilibrium independent of the initial state \cite{Diehl08} given by
the {\em pure many body state} $\rho_{ss}=|{\rm BEC}\rangle\langle
{\rm BEC}|$ representing a Bose Einstein condensate. From an atomic
physics point of view this is remarkable, as typical decoherence
mechanisms, such as spontaneous emission acting locally on lattice
sites, will destroy long range order, whereas here the bath coupling
is engineered to suppress phase fluctuations. This can be easily
understood in momentum space, where the annihilation part of
$c_{\ell\ell'}$ reads $ \sum_\lambda (1 - \exp(\mathrm i
\textbf{q}_\lambda a))b_\textbf{q}$, with $\lambda$ the reciprocal
lattice directions and $a$ the lattice constant. $c_{\ell\ell'}$ thus
feature a (unique) dissipative zero mode at ${\bf q} = 0$ -- a
many-body ``dark state'' $|\mathrm{BEC}\rangle\sim
b_{\textbf{q}=0}^{\dagger\, N}|{\rm vac}\rangle$ decoupled from the
bath, into which the system is consequently driven for long wait
times. The dynamics behind Eq.~(\ref{mastereq}) can thus be understood
as a ``dark state laser cooling" \cite{LaserCooling} into a
condensate, although in a many-body context.

\begin{figure}[t]
\includegraphics[width=\linewidth]{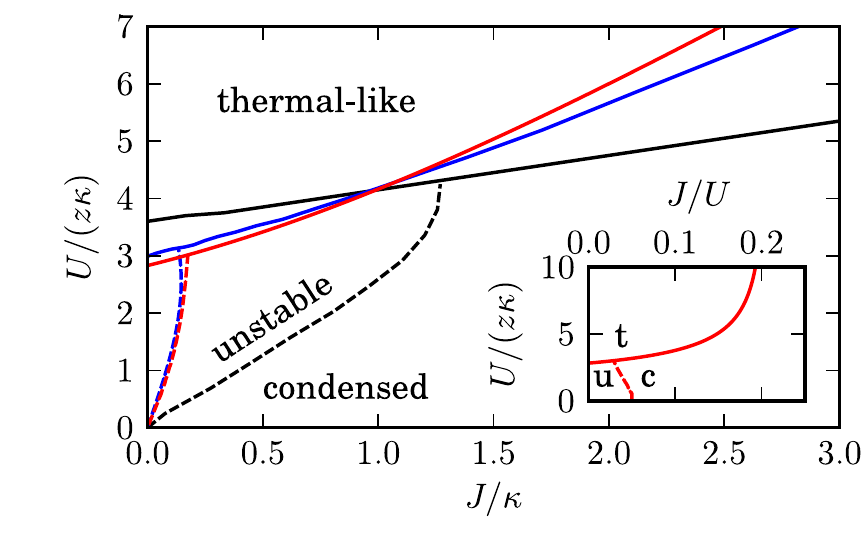}
\vspace{-0.7cm}
\caption{\label{fig:phasediagram} (color online) Nonequilibrium phase
  diagram for the model in Eq.~(\ref{eq:redmasterequation}).  The
  solid lines indicate the border of the dynamical quantum phase
  transition from a condensed to a homogeneous thermal steady state.
  The dashed lines delimit the region where the condensed state is
  stable with respect to spatial fluctuations.  The black (blue) lines
  are the numerical results corresponding to average density $n=1.0$
  ($n=0.1$).  The red line corresponds to the analytical results for
  $n=0.1$.  }
\end{figure}

$|\mathrm{BEC}\rangle$ is also an eigenstate of kinetic energy.  In
contrast, turning on an interaction measured by $u = U/(4\kappa z)$
provides a Hamiltonian term in (\ref{mastereq}) which is incompatible
with kinetic energy and dissipation. This competition leads to novel
dynamical equilibria which cannot be understood as thermodynamic
equilibrium states found from minimizing a free energy. They are
summarized in the steady state phase diagram in
Fig. \ref{fig:phasediagram}. Most prominently, it features a strong
coupling phase transition as a function of $u$. A first hallmark of
the nonequilibrium nature of the system is this: The transition shares
features of a QPT in that it is interaction driven, and of a classical
phase transition in that the ordered phase terminates in a mixed
state. This contrasts e.g. the well-known dissipation induced phase
transition to a superconductor in Josephson junction arrays
\cite{JJarrays}, in which detailed balance guarantees that the
system's state remains pure despite the suppression of phase
fluctuations via the coupling to a zero temperature bath.

Furthermore, we show the existence of a novel dynamical instability
that covers an extensive domain of the phase diagram. Again, this is a
nonequilibrium effect, since in equilibrium, finite momentum
excitations carry positive kinetic energy ruling out dynamical
instabilities.  It persists at arbitrarily weak interaction parameters
$Un$ due to its fluctuation induced nature elucidated below. This is
in marked contrast to the ``classical'' dynamical instabilities of
condensates in boosted lattices \cite{DynInstabTh,DynInstabExp} or in
exciton-polariton systems \cite{Carusotto10}, which are induced by
external tuning of parameters beyond finite critical values.

\emph{Nonlinear mean field master equation}.---To solve the master
equation we developed a generalized Gutzwiller approach, expected to
hold in sufficiently high spatial dimension, which allows to include
density matrices corresponding to mixed states.  This is implemented
by a product ansatz $\rho = \bigotimes_{\ell}\rho_{\ell}$, with the
reduced local density operators $\rho_{\ell} = \mathrm{Tr}_{\ne
  \ell}\,\rho$.  The equation of motion (EoM) reads
\begin{equation}\label{eq:redmasterequation}
\partial_{t}\rho_{\ell}  = -i [h_{\ell},\rho_{\ell}] +{\cal L}_{\ell}[\rho_{\ell}]~,
\end{equation}
with the local Hamiltonian $h_{\ell} = - J \sum_{\langle \ell' | \ell
  \rangle} (\langle b_{\ell'} \rangle b_{\ell}^{\dag} + \langle
b_{\ell'}^{\dag}\rangle b_{\ell} ) -\mu \hat{n}_{\ell} +\frac{1}{2} U
\hat{n}_{\ell}(\hat{n}_{\ell} - 1)$ reproducing the standard form of
the Gutzwiller mean field approximation and a Liouvillian of the form
$ {\cal L}_{\ell}[\rho_{\ell}] = \kappa \sum_{\langle \ell'| \ell
  \rangle} \sum_{r,s=1}^{4} \Gamma_{\ell'}^{rs}[2 A_{\ell}^{r}
\rho_{\ell} A_{\ell}^{s\dag} - A_{\ell}^{s\dag} A_{\ell}^{r}
\rho_{\ell} - \rho_{\ell} A_{\ell}^{s\dag} A_{\ell}^{r}]$.  The
Liouvillian is constructed with the vector of operators ${\bf A}_\ell
= (1, b_{\ell}^{\dag}, b_{\ell}, \hat{n}_{\ell})$ and the matrix of
correlation functions $\Gamma_{\ell}^{r,s} = \sigma^{r} \sigma^{s}
{\rm Tr}_{\ell} A_{\ell}^{(5-s)\dag} A_{\ell}^{(5-r)}\rho_{\ell}$, for
$\sigma = (-1,-1,1,1)$.  The $\rho$-dependent correlation matrix makes
the master equation \emph{nonlinear} in $\rho_\ell$.

\emph{Dynamical quantum phase transition}.---At $U=0$ a steady state
solution of Eq.~(\ref{eq:redmasterequation}) is given by the pure
state $\rho^{\rm (c)}_{\ell} = |\Psi\rangle\langle\Psi|$ for any
$\ell$ together with the choice $\mu = - Jz$, where $|\Psi\rangle$ is
a coherent state of parameter $n e^{i\theta}$ for any phase $\theta$
\cite{GardinerZoller}.  In order to understand the effect of a finite
interaction $U$, we apply the rotating-frame transformation
$\hat{V}(U) = \exp[i U \hat{n}_{\ell}(\hat{n}_{\ell} -1)t]$ to
Eq.~\eqref{eq:redmasterequation}.  This removes the interaction term
from the unitary evolution, but the annihilation operators become
$\hat{V} b_{\ell} \hat{V}^{-1} = \sum_{m} \exp(i m U t)
|m\rangle_{\ell}\langle m| b_{\ell}$.  The effect of a finite $U$ is
thus to rotate the phase of each Fock state differently, leading to
dephasing of the coherent state $\rho_{\ell}^{(c)}$. Hence, for strong
enough $U$, off-diagonal order is suppressed completely and the
density matrix becomes diagonal. In this case
Eq.~(\ref{eq:redmasterequation}) reduces precisely to the master
equation for a system of bosons coupled to a thermal reservoir with
occupation $n$~\cite{GardinerZoller}, whose solution is a mixed
diagonal thermal state $\rho^{\rm (t)}$.  Interestingly, this state is
thermal-like; however the role of the thermal bath is played by the
system itself, being provided by the mean occupation of neighbouring
sites.

\begin{figure}[t]
\includegraphics[width=\linewidth]{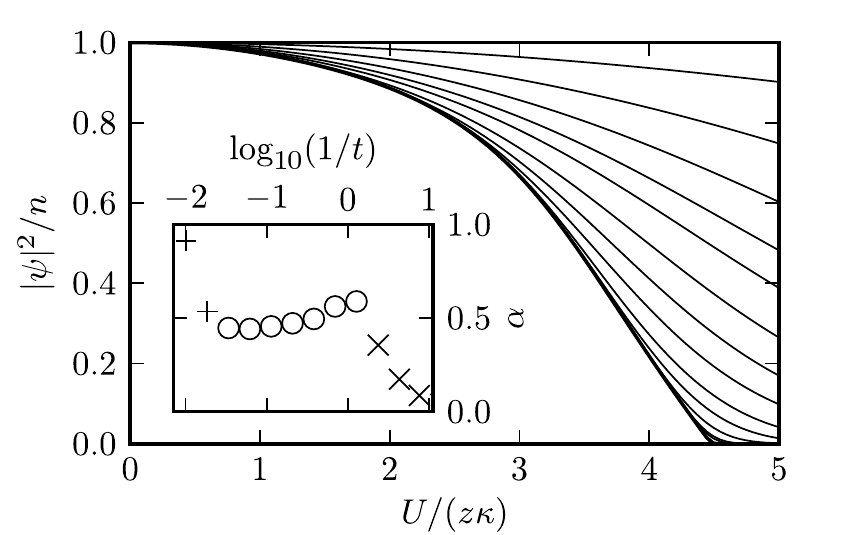}
\caption{\label{fig:timeconvergence} Stroboscopic plot of the time
  evolution of the condensate fraction as a function of the
  interaction strength $U$, for $J = 1.5\,\kappa$ and $n=1$. For large
  times it converges to the lower thick solid line.  The critical
  point is $U_{\rm c} \simeq 4.5\, \kappa z$.  Inset: Near critical
  evolution reflected by the logarithmic derivative of the order
  parameter $\psi(t)$, for $J=0$, $n=1$, and $U\lesssim U_{\rm c}$.
  The early exponential decay ($\times$) is followed by a scaling
  regime ($\circ$) with exponent $\alpha \simeq 0.5$. The final exponential
  runaway ($+$) is due to a small deviation from the critical point.
}
\end{figure}

We substantiate the discussion above with the numerical integration of
the EoM (\ref{eq:redmasterequation}) for a homogeneous system (we drop
the index $\ell$).  The system is initially in the coherent state and
the condensate fraction $|\psi|^{2}/n$, where $\psi = \langle b
\rangle$, decreases in time depending on the value of the interaction
strength $U$.  The result is a continuous transition from the coherent
state $\rho^{\rm (c)}$ to the thermal state $\rho^{\rm (t)}$, shown in
Fig.~\ref{fig:timeconvergence} for some typical parameters.  The
boundary between the thermal and the condensed phase with varying
$J,n$ is shown in Fig.~\ref{fig:phasediagram} with solid lines.

The transition is a smooth crossover for any finite time, but for
$t\to \infty$ a sharp nonanalytic point indicating a second order
phase transition develops.  In the universal vicinity of the critical
point, $1/\kappa t$ may be viewed as an irrelevant coupling in the
sense of the renormalization group. We may use this attractive
irrelevant direction to extract the critical exponent $\alpha$ for the
order parameter from the scaling solution $|\psi(t)| \propto (\kappa
t)^{-\alpha}$. In the inset of Fig.~\ref{fig:timeconvergence} we plot
$\alpha(t) = d \log (\psi) / d \log (1/t)$ and read off the critical
exponent $\alpha = 0.5$ in the scaling regime, which is an expected
result given the mean field nature of the Gutzwiller ansatz.  We
emphasize that following the relaxation dynamics of the condensate
fraction for critical system parameters gives an experimental handle
for the measurement of $\alpha$.

\emph{Low-density limit}.---In the low density limit $n\ll 1$ we
obtain an analytical understanding of the time evolution based on the
observation that the six correlation functions $\psi$, $\langle
b_{\ell}^{2} \rangle$, $\langle b_{\ell}^{\dag} b^{2}_\ell \rangle$,
and complex conjugates, form a closed (nonlinear) subset which
decouples from the \emph{a priori} infinite hierarchy of normal
ordered correlation functions $\langle b_{\ell}^{\dag n} b_{\ell}^{m}
\rangle$.  We first use this result to obtain analytically the
critical exponent $\alpha$ discussed above.  For a homogeneous system
with $J=0$ the EoMs read
\begin{eqnarray}\label{eq:corrfunccrit}
\partial_{t} \psi & = & i\mu \psi + (-iU + 4\kappa) \langle b^{\dag} b^{2} \rangle - 4 \kappa \psi^{\ast} \langle b^{2} \rangle ~, \nonumber \\
\partial_{t} \langle b^{\dag} b^{2} \rangle & = & 8 n \kappa \psi + (-iU + i \mu - 8 \kappa) \langle b^{\dag} b^{2} \rangle ~,\nonumber \\
\partial_{t} \langle b^{2} \rangle & = & (-iU + 2 i \mu - 8 \kappa) \langle b^{2} \rangle + 8 \kappa \psi^{2}~.
\end{eqnarray}
The structure of the equations suggest that $\langle b^{2} \rangle$
decays much faster than the other correlations for $U=U_\text{c}$, so
that we may take $\partial_{t} \langle b^{2} \rangle = 0 $ and hence
$\langle b^{2} \rangle \propto \psi^{2}$.  At the critical point the
two linear contributions to $\partial_t \psi$ vanish due to the zero
mass eigenvalue at criticality and $\partial_{t} \psi \propto \kappa
\psi^{2}\psi^{\ast}$.  It follows that $|\psi| \simeq 1 / (4
\sqrt{\kappa t})$ in agreement with the numerical result in
Fig.~\ref{fig:timeconvergence}.

To study the interaction induced depletion of the condensate fraction,
it is convenient to use ``connected'' correlation functions, built
with the fluctuation operator $\delta b = b - \psi_{0}$.  Here
$\psi_{0}$ is the constant value of the order parameter in the steady
state, and $\langle \delta b \rangle = 0$. From
(\ref{eq:corrfunccrit}) we obtain a closed \emph{linear} system of
EoMs, if $\psi_0$ is considered as a parameter, determined
self-consistently from the identity $n = \langle \delta b^{\dag}
\delta b \rangle + |\psi_{0}|^{2}$.  The value of the chemical
potential is fixed to remove the driving terms in the equations for
$\langle \delta b \rangle $, leading to $\mu = n U$.  This is an
equilibrium condition similar to the vanishing of the mass of the
Goldstone mode in a thermodynamic equilibrium system with spontaneous
symmetry breaking.  The solution of the equations in steady state
yields the condensate fraction
\begin{equation}\label{eq:Depletion}
\frac{|\psi_0|^{2}}{n} = 1 - \frac{2 u^2 \left(1+(j + u)^2\right)}{1+u^2 + j( 8 u +6 j \left(1+2 u^2\right) +24 j^2 u + 8 j^3)}~,
\end{equation}
with dimensionless variable $j = J/(4\kappa)$.
Eq.~\eqref{eq:Depletion} reduces to the simple quadratic expression $1
- 2u^2$ in the limit of zero hopping, with the critical point
$U_{c}(J=0) = 4 \kappa z/\sqrt{2}$.  The phase boundary, obtained by
setting $\psi_{0} = 0$ in Eq.~\eqref{eq:Depletion}, reads $u_c =
j+\sqrt{ 1 / 2 + 2 j^2}$.  Fig.~\ref{fig:phasediagram} shows that
these compact analytical results (solid red line) match the full
numerics for small densities (solid blue line), and also explain the
qualitative features of the phase boundary for large densities. We
note the absence of distinct commensurability effects for e.g. $n=1$,
tied to the fact that the interaction also plays the role of heating.

\emph{Dynamical instability}.---Numerically integrating the full EoM
(\ref{eq:redmasterequation}) with site-dependence (in one dimension
for simplicity), we observe a dynamical instability, manifesting
itself at late times in a long wavelength density wave with growing
amplitude.  Numerical linearization
of Eq.~(\ref{eq:redmasterequation}) around the homogeneous steady
state allows to draw a phase border for the unstable phase (see
Fig.~\ref{fig:phasediagram}).  The instability is cured by the
increase of hopping $J$, which is associated to an operator compatible
with dissipation $\kappa$.  Furthermore, we note that the thermal
state is always dynamically stable against long wavelength
perturbations.

\begin{figure}[t]
\includegraphics[width=\linewidth]{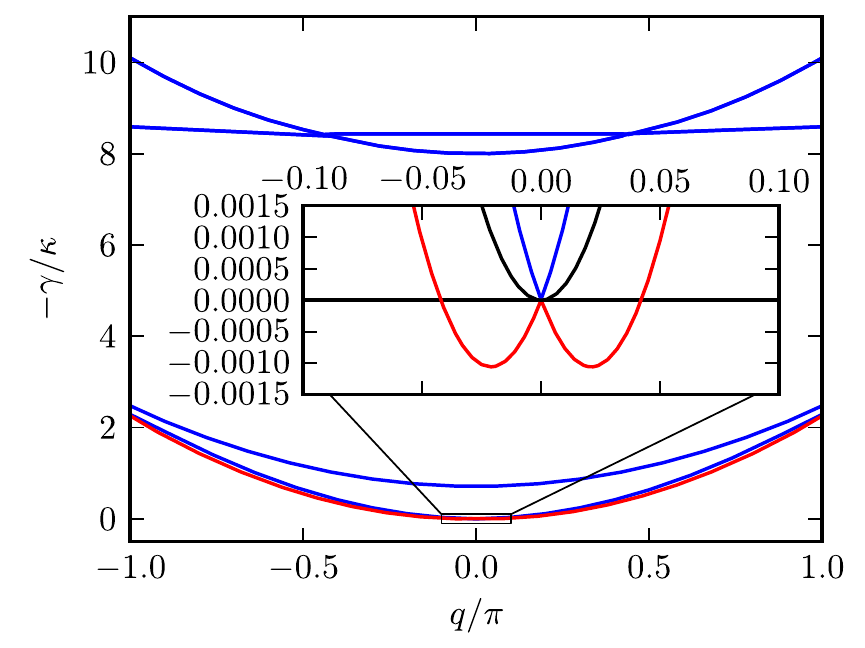}
\caption{\label{fig:spectrum} Real (dissipative) part of the spectrum
  $\gamma_\textbf{q}$ from the analytical low density limit for $J=0$,
  $n=0.1$, and $U = 1.0\,\kappa$. The inset magnifies the parameter
  region with unstable modes (red solid line).  The black solid line
  is the bare dissipative spectrum $\kappa_{\textbf{q}}$. }
\end{figure}

The origin of this instability is intriguing and we discuss it
analytically within the low-density limit introduced above.  We
linearize in time the EoM (\ref{eq:redmasterequation}), writing the
generic connected correlation function as $\langle \hat{\cal O}_{\ell}
\rangle(t) = \langle \hat{\cal O}_{\ell} \rangle_{0} + \delta \langle
\hat{\cal O}_{\ell} \rangle(t)$, where $\langle \hat{\cal O}_{\ell}
\rangle_{0}$ is evaluated on the homogeneous steady state of the
system.  The EoM for the time and space dependent fluctuations is then
Fourier transformed, resulting in a $7\times7$ matrix evolution
equation $\partial_{t} \delta \Phi_{\textbf{q}} = M \delta
\Phi_{\textbf{q}}$ for the correlation functions $\Phi_{\textbf{q}} =
(\langle \delta b\rangle_{\textbf{q}},$$ \langle \delta
b^{\dag}\rangle_{\textbf{q}}, \langle \delta b^{\dag} \delta b
\rangle_{\textbf{q}},$$ \langle \delta b^{2} \rangle_{\textbf{q}},$$
\langle \delta b^{\dag 2} \rangle_{\textbf{q}}, \langle \delta
b^{\dag} \delta b^{2} \rangle_{\textbf{q}},$$ \langle \delta b^{\dag
  2} \delta b \rangle_{\textbf{q}} )$.  We note that the fluctuation
$\delta \langle \delta b \rangle_{\textbf{q}}$ ($\delta \langle \delta
b^{\dag} \rangle_{\textbf{q}}$) coincides with the fluctuation of the
order parameter $\delta \psi_{\textbf{q}}$ ($\delta
\psi_{-\textbf{q}}^{\ast}$).  The full matrix $M$ can be easily
diagonalized numerically revealing the spectrum in
Fig.~\ref{fig:spectrum} (we display only the relevant real part
$\gamma$ corresponding to damping).  The lowest-lying branch gives
$\gamma_\textbf{q}< 0$ in a small interval around the origin
$\textbf{q}=0$.  This means that the correlation functions grow
exponentially $\propto e^{\gamma t}$ in a range of low momenta,
resulting e.g.~in a long wavelength density wave.

Due to the scale separation for $\textbf{q} \to 0$ in the matrix $M$
apparent from Fig.~\ref{fig:spectrum}, we can apply second order
perturbation theory twice in a row to integrate out the fast modes
$\gamma \propto \kappa$ and $ \propto \kappa n$. We then obtain an
effective low energy EoM for the fluctuations of the order parameter
$(\delta \psi_{\textbf{q}} , \delta \psi_{-\textbf{q}}^{\ast})$,
governed by a $2\times 2$ matrix
\begin{eqnarray}\label{eq:linear2x2}
M_{\text{eff}} = \left(
\begin{array}{cc}
Un + \epsilon_{\textbf{q}} - i \kappa_{\textbf{q}}  &  Un + 9  u n \kappa_{\textbf{q}}   \\
 - Un -  9 u n \kappa_{\textbf{q}}  &  - Un -  \epsilon_{\textbf{q}}  - i \kappa_{\textbf{q}}   
\end{array}
\right)~,
\end{eqnarray}
where $\epsilon_{\textbf{q}}=J\textbf{q}^{2}$ represents the kinetic
contribution and $\kappa_{\textbf{q}}=2(2n+1)\kappa \textbf{q}^{2}$ is
the bare dissipative spectrum.  The form of the EoM reflects the
structure of the spatial fluctuations which are included in our
approach, that may be understood as scattering off the mean fields in
opposite directions.  We note that a naive \emph{a priori} restriction
to the $2\times2$ set corresponding to the subset
$(\delta\psi_{\ell},\delta\psi_\ell^{\ast})$ would be inconsistent,
for example destroying the dark state property present in the correct
solution $M_{\text{eff}}$.  On the other hand, factorizing the
correlation functions in the Liouvillian ${\cal L}_{\ell}$ yields a
dissipative Gross-Pitaevski equation but its linearization in time
produces a matrix $M_{\text{eff}}$ \emph{without} the fluctuation
induced term $\sim u$ and fails to describe the dynamical instability.
Thus, in order to correctly capture the physics of the instability at
long wavelength $\textbf{q}\to 0$, the onsite quantum correlations
renormalizing $M_{\text{eff}}$ have to be properly taken into account.

We can make the nature of the instability even more transparent from
calculating the lowest eigenvalue of $M_{\text{eff}}$,
$\gamma_\textbf{q} \simeq \mathrm i c |\textbf{q}| +
\kappa_{\textbf{q}}$, with speed of sound $c=\sqrt{2 U n[J - 9 U n /
  (2z) ]}$. If the hopping amplitude is smaller than the critical
value $J_{\rm c} = 9Un / (2z)$ the speed of sound turns imaginary and
contributes to the dissipative real part of $\gamma_\textbf{q}$. The
nonanalytic renormalization contribution $\sim |\textbf{q}|$ always
dominates the bare quadratic piece for low momenta, explaining the
shape in the inset of Fig.~\ref{fig:spectrum} and rendering the system
unstable.  The linear slope of the stability border for small $J$ and
$U$ is clearly visible from the numerical results in
Fig.~\ref{fig:phasediagram}.  In summary, the origin of the
instability is traced back to a subtle renormalization effect of the
speed of sound at low energies, which in turn is due to an interplay
of short time quantum and long wavelength classical fluctuations.

\emph{Conclusion}.---We have discussed the steady state phase diagram
resulting from a competition of unitary Bose-Hubbard and dissipative
dynamics with dark state. The features found in the present model are
expected to be generic and representative for a whole class of
nonequilibrium models discussed recently in the context of reservoir
engineering and dissipative preparation of given long range ordered
entangled states of qubits or spins on a
lattice~\cite{Verstraete09,Weimer10} and paired fermions
\cite{Diehl08,Yi10}. In particular, we emphasize the importance of a
compatible energy term for the achievement of stability of
driven-dissipative many-body systems in future experiments.

\acknowledgements We thank M. Hayn, A. Pelster, S. Kehrein,
M. M\"ockel, and J. V. Porto for interesting discussions.  This work
was supported by the Austrian Science Foundation through SFB FOQUS,
SCALA and by EU Networks.

\end{document}